\documentclass[pdflatex,sn-mathphys-num,twocolumn]{sn-jnl}

\usepackage{graphicx}%
\usepackage{multirow}%
\usepackage{amsmath,amssymb,amsfonts}%
\usepackage{amsthm}%
\usepackage{mathrsfs}%
\usepackage[title]{appendix}%
\usepackage{xcolor}%
\usepackage{tcolorbox}
\tcbuselibrary{breakable}
\usepackage{ulem}
\usepackage{textcomp}%
\usepackage{manyfoot}%
\usepackage{booktabs}%
\usepackage{algorithm}%
\usepackage{algorithmicx}%
\usepackage{algpseudocode}%
\usepackage{listings}%
\usepackage{longtable}
\def\BibTeX{{\rm B\kern-.05em{\sc i\kern-.025em b}\kern-.08em
    T\kern-.1667em\lower.7ex\hbox{E}\kern-.125emX}}
\usepackage{colortbl}
\usepackage{hyperref}
\usepackage{verbatim}
\hypersetup{
    colorlinks=true,
    linkcolor=blue,
    filecolor=magenta,      
    urlcolor=cyan,
    }
\usepackage{float}
\newfloat{algorithm}{t}{lop}

\algrenewcommand\algorithmicrequire{\textbf{Input:}}
\algrenewcommand\algorithmicensure{\textbf{Output:}}

\newcommand{\eat}[1]{}

\theoremstyle{thmstyleone}%
\theoremstyle{thmstyletwo}%

\theoremstyle{thmstylethree}%

\begin{document}

\title[Social Media Data Mining of Human Behaviour during Bushfire Evacuation]{Social Media Data Mining of Human Behaviour during Bushfire Evacuation}


\author[1]{\fnm{Junfeng} \sur{Wu}}\email{wujunfeng@vip.163.com}

\author*[1]{\fnm{Xiangmin} \sur{Zhou}}\email{xiangmin.zhou@rmit.edu.au}

\author[1]{\fnm{Erica} \sur{Kuligowski}}\email{erica.kuligowski@rmit.edu.au}

\author[2]{\fnm{Dhirendra} \sur{Singh}}\email{sin122@csiro.au}

\author[3]{\fnm{Enrico} \sur{Ronchi}}\email{enrico.ronchi@brand.lth.se}

\author[4]{\fnm{Max} \sur{Kinateder}}\email{Max.Kinateder@nrc-cnrc.gc.ca}

\affil*[1]{\orgdiv{Department}, \orgname{RMIT University}, \orgaddress{\street{G.P.O. Box 2476}, \city{Melbourne}, \postcode{3001}, \state{Victoria}, \country{Australia}}}

\affil[2]{\orgdiv{Department}, \orgname{CSIRO}, \orgaddress{\street{Private Bag 10 Clayton South VIC 3169}, \city{Melbourne}, \postcode{3169}, \state{Victoria}, \country{Australia}}}

\affil[3]{\orgdiv{Department}, \orgname{Lund University}, \orgaddress{\street{P.O. Box 118}, \city{Lund}, \postcode{22100}, \country{Sweden}}}

\affil[4]{\orgdiv{Department}, \orgname{National Research Council Canada}, \orgaddress{\street{1200 Montreal Road}, \city{Ottawa}, \postcode{K1A 0R6}, \state{Ontario}, \country{Canada}}}


\abstract{Traditional data sources on bushfire evacuation behaviour, such as quantitative surveys and manual observations have severe limitations. Mining social media data related to bushfire evacuations promises to close this gap by allowing the collection and processing of a large amounts of behavioural data, which are low-cost, accurate, possibly including location information and rich contextual information. However, social media data have many limitations, such as being scattered, incomplete, informal, etc.
Together, these limitations represent several challenges to their usefulness to better understand bushfire evacuation. To overcome these challenges and provide guidance on which and how social media data can be used, this scoping review of the literature reports on recent advances in relevant data mining techniques. In addition, future applications and open problems are discussed. We envision future applications such as evacuation model calibration and validation, emergency communication, personalised evacuation training, and resource allocation for evacuation preparedness. We identify open problems such as data quality, bias and representativeness, geolocation accuracy, contextual understanding, crisis-specific lexicon and semantics, and multimodal data interpretation.}




\maketitle

\section{Introduction \label{sec:introduction}}
Bushfires\footnote{(often large) unplanned fires the wildland outside urban, industrial, or other infrastucture; also referred to as wildfires} have caused a large number of fatalities in Wildland Urban Interface (WUI) and rural areas around the world \cite{Rodney2021,Blanchi2014192}.
As the climate changes, global temperatures are increasing and droughts are becoming more common around the world, thus bushfires are likely to occur more frequently and intensely \cite{Jolly2015,sawyer2020tip}.
Bushfires can cause even more damage with poor land management and increased urbanisation of the WUI \cite{Koksal201921}.
Therefore, bushfire evacuation planning is more vital than ever \cite{Kuligowski2021}.

Evacuation behaviours of people in any hazardous event can be described as a sequence of decisions and protective actions \cite{Kuligowski2021}. People may show a wide variability of decisions and actions, even under the same circumstances. Even the same people may react differently in the same scenario. The most consequential decision to make is whether to leave, stay, or wait and see. For people who decide to leave, there are subsequent decisions that involve departure time (i.e., when to leave one's starting location), mode of transportation (e.g., via a private vehicle), as well as destination (where to evacuate to) and route choice (the path to reach a place of safety). The departure time can vary significantly, depending on when people receive an evacuation order and their level of preparedness. For example, while some may decide to evacuate early (shadow evacuation), others may not be prepared or initially decide to stay but then change their mind \cite{Ronchi20141545}. Those who decide to stay have to make other decisions, such as whether to defend actively or inactively \cite{whittaker2017119,Whittaker2020} or where to find a safe shelter. These decisions are affected by many factors, including sociodemographic factors, environmental factors, social cues, experience, level of preparedness, societal responsibilities, location, and perceived risk \cite{Folk20191619}.


The focus of this scoping review is the analysis of
human behaviour for bushfire evacuation planning.
Evacuation planning is the process of preparing and facilitating the safe movement of people away from
the threat of a bushfire. 
Effective evacuation planning requires a deep understanding of these complex, variable human behaviors under threat. It involves identifying risks, designing strategies (e.g. trigger buffers \cite{Mitchell2023}), and optimising resources for the evacuation process. However, obtaining data of sufficient quality, scale and realism to inform such plans is a significant challenge. Recently, the popularity of online social services has provided a vital data source to mining bushfire evacuation behaviors, enabling effective and efficient disaster management.  



Unlike traditional quantitative surveys
\cite{Toledo20181366,Katzilieris2022} and manual observations \cite{Gwynne2023879} of evacuation behaviors that are scenario-specific and limited in sample size and composition \cite{Smyth20101423}, social media data appear on online platforms in real time with large behavioral data during disasters.
These platforms are frequently used by both affected residents and emergency officials for communication and information sharing during crises \cite{Bennett2017303}. For example, residents may post about their intentions, actions, or needs, while officials disseminate warnings and orders. Many of these activities contain traces of users' physical activities and decision-making processes \cite{Wukich2023}. Unlike retrospective surveys, social media data are often produced in real-time, offering a less structured but more immediate record of human response.
Because of these advantages, social media data have been used in various case studies of natural hazards to study human behavior, such as hazardous event detection  \cite{Sakaki2010851}, hurricanes evacuation behaviors \cite{Martín2020}, and bushfire evacuation decision making \cite{Nara2017190}. These studies demonstrate the potential of social media data to improve pre-event evacuation planning and guidance. However, extracting useful information from social media for bushfire evacuation planning presents several challenges:


\begin{itemize}
\item {Overgeneralization}: social media platforms are
general-purpose tools which are not necessarily
designed for emergencies.
\item {Scatteredness}: information is scattered across multiple platforms, users, and intended audiences.
\item {Incompleteness}: posts often have character limits or design constraints, leading to incomplete information. Privacy settings often mean location data is unavailable or imprecise \cite{Martí2019161} 
\item {Informality}: posts on social media are written in an informal language with many acronyms, abbreviations,
and emojis that may be hard to interpret for researchers and users \cite{was2021did}.
\item {Implicitness}: useful information is often implicit, hidden in conversations, and requires the understanding of local context.
\item {Bias}: data can be biased towards areas with better internet access and younger demographics \cite{Martí2019161}.
\item {Hyperlocal data}: data from one region often cannot be directly transferred to analyze another due to cultural and demographic differences.
\item {Large volume}: the flow of data is immense and requires efficient processing techniques.
\item {Low signal-to-noise ratio}: only a small fraction of posts are relevant for hazard and evacuation \cite{Martín2017}. 
\end{itemize}

These challenges necessitate the use of specialized data mining and machine learning techniques to process and analyze social media data effectively \cite{Kumar2020,ochoa2021machine}. This motivates the need for a scoping review to map and assess the techniques that can be used to overcome these limitations for the specific purpose of understanding bushfire evacuation behavior.

\noindent\textbf{Previous reviews and the current scoping review.}
Previous reviews have laid important groundwork. Houston, J.B. et al. \cite{Houston20151} developed a framework for social media use in emergencies, and Zhang C. et al. \cite{Zhang2019190} envisioned intelligent systems for disaster guidance. Others have focused on specific hazards like hurricanes \cite{Martín2020} or presented data-driven case studies for wildfires \cite{Li2021, morshed2021trend}. While these works are valuable, a systematic review of the data mining techniques themselves—especially as applied to the unique context of bushfires and with a focus on informing computational evacuation models—is lacking. This scoping review aims to fill that gap by specifically focusing on the methods used to mine social media data for insights into bushfire evacuation behavior.

    

\noindent\textbf{Method of scoping review.}
This study was conducted as a systematic scoping review following established guidelines for transparent reporting in research synthesis. The methodology was designed to ensure comprehensive coverage of relevant literature while maintaining rigorous selection standards.

The research process began with systematic searches across three major academic databases: Scopus, Web of Science, and Google Scholar, conducted between May and October 2023. Search terms were carefully constructed to capture studies at the intersection of social media analysis and bushfire evacuation behavior, combining platform references, hazard terminology, behavioral focus, and analytical methods.

Initial searches returned 329 potentially relevant papers. After removing 42 duplicate records, we implemented a structured screening approach. The first screening phase evaluated titles and abstracts against predetermined criteria. Recognizing the nascent stage of bushfire-specific social media research, our initial criteria were broadened to include studies on other rapid-onset hazards (e.g., hurricanes, earthquakes) to capture methodologically relevant approaches that could be transferable to bushfire contexts. Papers were excluded if they: (a) did not focus on rapid-onset hazard evacuation, (b) did not use social media as a primary data source, (c) were not peer-reviewed, or (d) lacked a clear methodology. This resulted in the exclusion of 94 papers.

The remaining 193 papers advanced to full text review, including 117 journal papers, 67 conference / workshop papers, 8 books / book chapters, and 1 technical report. During this phase, each study was thoroughly examined to extract key information on research objectives, methodological approaches, behavioral findings, and study limitations. Particular attention was paid to how different studies operationalized social media data collection and analysis in the context of bushfire evacuations.

The selection process emphasized studies providing actionable insights into evacuation behaviors while representing various methodological methods appropriate for a scoping review. Throughout the process, we maintain detailed documentation of inclusion and exclusion decisions to ensure the reproducibility of our methodology. This allowed us to systematically map the current state of research while identifying important gaps in the literature.

The methodology was designed to address several key challenges in this research area, including the interdisciplinary nature of the topic, varying terminologies across fields, and the need to balance breadth with depth of coverage. By combining systematic search procedures with careful screening and documentation, we ensured the review's findings are comprehensive and methodologically sound.    

\noindent\textbf{Structure and contributions of the present study.}
In this study, we review existing papers to give an overview of the concepts, techniques, and future applications for social media data mining of human behaviour during bushfire evacuations. All concepts in Section \ref{sec:overview} are directly related to understanding evacuation behaviour. The techniques are investigated in Section \ref{sec:tech} for overcoming the challenges mentioned above. We discuss future applications in Section \ref{sec:applications} and open problems in Section \ref{sec:problems} that require future study. In pushing forward this research, we hope to highlight the important concepts and techniques used to analyse social media data to obtain a better understanding of evacuation behaviour in bushfires. This understanding (i.e.,  evacuation behaviour in communities/bushfire events including observed trends/patterns) can eventually be used in future applications to protect the lives of people living in fire-prone areas.  
The key contributions of this survey are three-fold: 
\begin{enumerate}
    \item We conduct a scoping review for the techniques of social media data mining of human behaviours during bushfire evacuations and propose a framework to the procedure of such data mining.
    \item We provide an overview and summary for selected future applications.
    \item We discuss the challenges and open problems in this research field.
\end{enumerate}

\section{Overview of social media data mining for bushfire evacuation \label{sec:overview}}
Social media data mining in this context aims to investigate bushfire evacuation behaviours and their timing and reasons. Despite uncertainty in general, human behaviour presents some degree of predictability with respect to a number of factors, including individual perceptions or available information. Due to such predictability, bushfire evacuation models can be subjected to testing to investigate their predictive capabilities \cite{Ronchi20231493}. 
Evacuation behaviour in bushfires consists of a series of decisions and actions. To design an effective evacuation plan, it is crucial to understand these decisions,  the time associated with them, and the time associated with the corresponding actions. There are many approaches to model evacuation behaviours (i.e. theory-driven or data-driven), but in this section, we focus on explaining the evacuation behaviours based on commonly adopted methods of traffic engineering. To present evacuation behaviours, we adopt the traditional four-step transport modelling approach \cite{Murray-Tuite201325,Pel201297,Intini2019}, which consists of the following four steps :
\begin{itemize}
    \item trip generation (i.e., to predict the number of evacuees and their departure time), 
    \item trip distribution (i.e., to predict the trip destination(s) or itineraries of evacuees),
    \item modal split (i.e., to predict evacuees’ modes of transport),
    \item trip assignment (i.e., to predict routes used by evacuees to reach the destination).
\end{itemize}
Although the four-step model is used in many non-emergency (i.e. routine) traffic models to understand travel behaviour, it can also be applied to understand evacuation behaviours in a bushfire context \cite{Intini2019}. Additionally, we consider these four steps as a useful starting point for a broader analysis of relevant information related to the understanding of evacuation behaviours, especially regarding their reasons and timing. 

\subsection{Trip Generation \label{ssec:trip_generation}}
Given information available on regional demography and behavioural factors that affect the evacuation decision \cite{McLennan2019487}, trip generation models predict the number of evacuees and their departure time \cite{Murray-Tuite201325,Pel201297}. Trip generation can be represented with several modelling approaches such as statistical models, descriptive models or machine learning. The difference between these types of models lies in how they handle the diversity of evacuation decisions and their overall predictive modelling approach. Statistical models such as \textcolor{black}{Random utility models} \cite{Murray-Tuite201325,Mozumder2008415,Sadri201750,Wong20231165} and other regression models \cite{Xu2023793,Katzilieris2022} diversify evacuation decisions by the probability of evacuation among $n$ alternative options. Descriptive models \cite{Ortúzar2011,Meyer20161} diversify decisions by classifying people in the evacuation region into several groups. This allows the prediction of different decisions for different groups. Machine learning models diversify decisions based on big data analysis of a large pool of factors that affect decisions and their timing. The factors are often automatically defined and extracted. Examples of previously applied machine learning approaches applied to this scope include naïve Bayes classifier, K-nearest neighbors, support vector machine, neural network, classification and regression tree (CART), random forest, and extreme gradient boosting.

Trip generation models generally represent multiple factors that affect evacuation decision making. For example, an individual's belief in their ability to act has been identified as a significant factor in some bushfire evacuation studies \cite{Strahan2019146,McCaffrey20181390}. Other examples include experience with evacuation \cite{Murray-Tuite201325}, types of evacuation instructions (that is, voluntary or mandatory) \cite{Mozumder2008415}, and the influence of social networks \cite{Sadri201750}. In addition to these examples, many other factors affecting bushfire evacuations were reviewed in \cite{Folk20191619,McLennan2019487}, such as environmental and social cues, experience and preparedness, familial and social responsibilities, location (e.g., proximity to the risk), credible threat and risk assessments, and regional evacuation policies. Therefore, it is of interest to obtain details on evacuation decisions (i.e., whether people stay or go), the reasons why they may have made that choice, and when they left their home (or original) location from social media messages posted during a fire event.

The choice of modelling approach involves a key trade-off. \textit{Traditional models} (statistical, descriptive) are often more interpretable and rely on well-established theory, but may struggle with complexity and high-dimensional data. \textit{Machine learning models} excel at capturing non-linear relationships from large datasets but can act as "black boxes" and risk overfitting to specific events. The suitability of each method depends on the research question, data availability, and the need for interpretability versus predictive power. Social media data, with its volume and unstructured nature, presents both an opportunity for ML applications and a challenge for traditional methods, a theme explored further in Section 3.
    

\subsection{Trip Distribution \label{ssec:trip_distribution}}
Given the data from the trip generation models, as well as the location of evacuation shelters and characteristics of the affected population, the trip distribution models predict the distribution of evacuees to different evacuation destinations. Trip generation models include those that predict final destinations \cite{Cuéllar2009628} and those that predict intermediate trips or stops (i.e. behavioural itineraries) \cite{Murray-Tuite2004150,Lindell200718}. Research has observed that the likelihood of destination options is mainly based on origin-destination (OD) travel costs and utilities, but decisions can be diversified according to individual and group factors. In particular, intermediate stop models, such as activity models \cite{Intini2019}, can produce trip chains using the O-D cost matrix \cite{MurrayTuite2003}. Random utility models \cite{Murray-Tuite201325,Wong2020331} and descriptive models \cite{Pel201297,Murray-Tuite2004150}) are also used to diversify destination decision-making.  Machine learning models can identify and interpret several factors and can make use of large datasets on destination choices \cite{Pourebrahim2019}. 

An example of destination choices is reported in an investigation of wildfire events in San Diego. The study reported that the destinations chosen during bushfires can include relative's home (43.6\%), friend's home (27.6\%), public shelter (4.9\%), hotel or motel (11\%), other (7.6\%) and original home locations (30. 4\%) \cite{sorensen2009effectiveness}, however these percentages can be different across fire events. 
\textcolor{black}{Wong et al 2023} \cite{Wong20231165} provided evidence of the heterogeneity of these percentages in response to different wildfires.
\textcolor{black}{Vaicuilyte et al 2022} \cite{Vaiciulyte2022} showed regional differences in such percentages.
Thus, it is of interest to obtain the following information from the analysis of social media during bushfires: type of destination \cite{Wong20231165-1}, factors that affect travel costs such as destination distances and route availability \cite{Cheng2011125}, and factors that influence decisions about destination choices \cite{Cheng2011125,Intini2018280}. More information on these factors and the types of destination can improve trip distribution modelling.

The modelling of trip distribution presents a trade-off between behavioural realism and computational complexity. \textit{Traditional models} (e.g., random utility, descriptive) provide a clear, theory-based understanding of destination choice factors but often rely on simplified assumptions that may not capture the full complexity of emergency decision-making. \textit{Machine learning approaches} can uncover subtle, non-linear patterns from large datasets (e.g., social media, GPS traces) but may lack interpretability and require very large volumes of data to generalize well across different fire events. The choice between them hinges on whether the research goal is explanation or prediction, a challenge that is central to the data mining techniques discussed in Section 3.

\subsection{Modal Split \label{ssec:modal_split}}
Given the data from the trip distribution, the data on transportation systems and the affected population, modal split modelling estimates the number of evacuees using different types of evacuation transportation modes, for example, private vehicles, public transportation, walking, or unconventional means of evacuation such as via sea or air  \cite{ronchi2023evacuation}. The main modal split approaches are descriptive approaches \cite{Sadri201437,Wong2020}, random utility approaches \cite{Wong20231165},  activity model approaches \cite{Xuwei201521,Lin200969} and machine learning approaches \cite{Zhao202022}. Descriptive approaches classify evacuees into several groups to produce different transport choices for different groups of evacuees. Random utility approaches randomise the choices according to probabilities. Activity model approaches generally simulate the process of choosing transport at the individual level according to each individual's specific situation. Machine learning approaches such as support vector machines or artificial neural networks have been used to interpret large datasets of variables related to modal split such as transport cost, residence, and demographics \cite{Omrani2015840}.  

The analysis of social media data can be important for our understanding of the modes people choose for evacuation. For example, \textcolor{black}{Toledo et al 2018} \cite{Toledo20181366} reported 92\% of evacuees from an Israeli bushfire evacuated using private vehicles, but this percentage is context-specific \cite{Wong20231165}. Social media data mining can extract context-specific information to discover the factors that influence these choices.

Selecting an approach for modal split analysis involves a trade-off between granularity and generalizability. \textit{Descriptive and random utility models} are powerful for testing specific hypotheses about how transport costs or demographics influence choices, but their effectiveness depends on pre-defined variables and categories. \textit{Activity-based and machine learning models} offer a more granular, individual-level view, better suited for simulating complex, interactive decision processes. However, they demand more detailed data and computational resources. The "best" method is context-dependent, influenced by data availability and the specific evacuation planning question at hand, a dilemma further explored in Section 3.

\subsection{Traffic Assignment \label{ssec:traffic_assignment}}
Given the data from the modal split, the data on transportation systems, and specific factors of the population, traffic assignment forecasts the routes that evacuees are likely to take to reach their destination. Prediction methods include static assignment methods and dynamic assignment methods. Static assignment methods  \cite{Hoogendoorn2010450} use a given origin-destination travel cost matrix (e.g. a at peak traffic) of the traffic network for prediction. Dynamic methods \cite{Pel201297} consider changes in traffic conditions over time and simulate the evolution of traffic conditions for prediction. The latter are recommended for use in evacuation applications given the possible dynamic availability of routes due to the wildfire spread \cite{Intini2019}.

Route choices are affected by many factors, such as preferences for more familiar routes over the shortest or quickest routes \cite{Murray-Tuite201298}, previous experience or route traffic conditions \cite{Wu2012445}, accessibility of the route, type of road (e.g., carriageway configuration), length of the route, and perceived service availability (e.g., gas stations located along the route) \cite{Dow200212}. The impact of these factors, and subsequent routing choices can be inferred via social media data mining, thus allowing understanding these behaviours in various fire and evacuation conditions. 

The choice between static and dynamic assignment methods underscores a fundamental trade-off between simplicity and realism. \textit{Static methods} are computationally efficient and simpler to implement but are ill-suited for evacuations, where network conditions and fire threats change rapidly. \textit{Dynamic methods} are essential for realistic evacuation modeling but are computationally intensive and require continuous, high-quality data on traffic and fire spread. Inferring route choice factors from social media could significantly enhance dynamic models, but this introduces challenges of data integration and real-time processing, which are key topics in Section 3.

\section{Relevant Techniques \label{sec:tech}}

\begin{figure*}[htb!]
    \centering
    \includegraphics[width=0.6\linewidth]{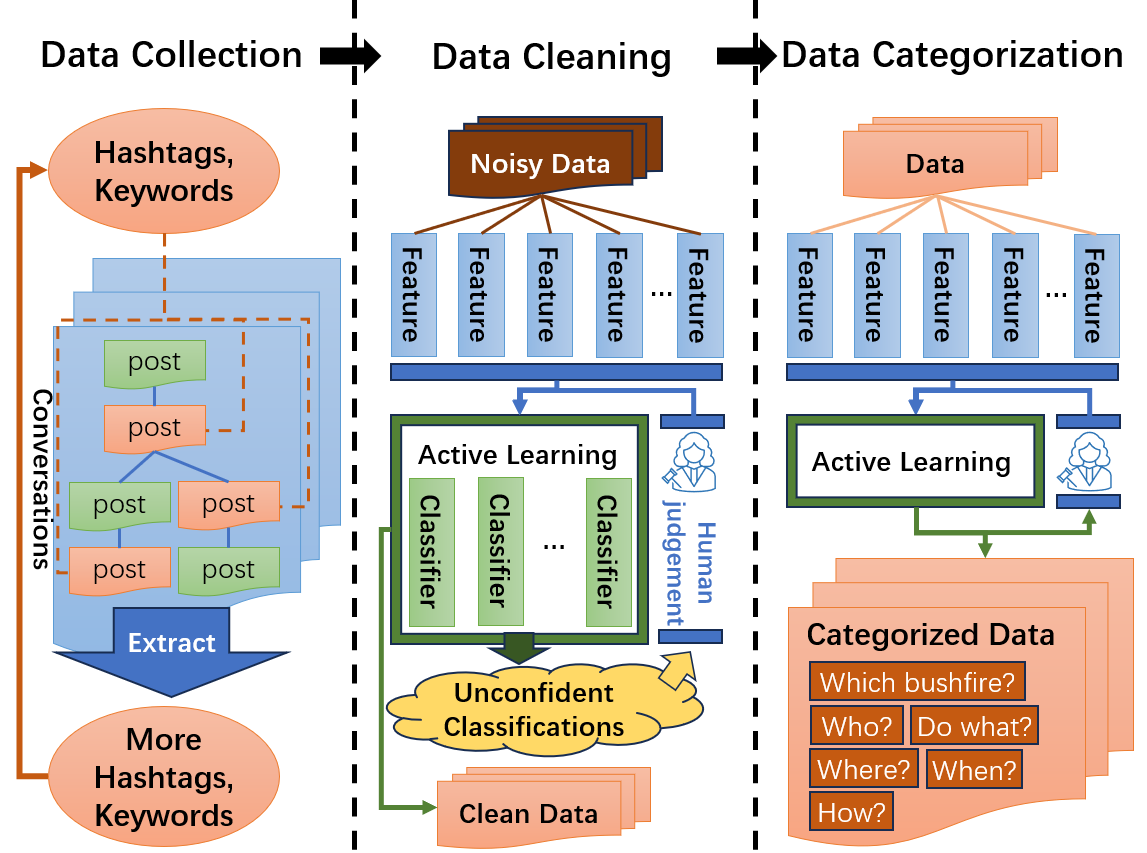}
    \caption{Pipeline of Social Media Data Mining of Human Behaviour during Bushfire Evacuations}\vspace{-2ex}
    \label{fig:pipeline}
\end{figure*}

This section reviews the scientific literature for the study of the techniques for social media data mining for bushfire evacuation. Recent wildfire evacuation studies \cite{Katzilieris2022,Xu2023793,2025Understanding} have demonstrated the growing importance of these techniques in understanding evacuation behaviors. 
In particular, we envision a pipeline of analysis that includes three stages: the collection, cleaning, and categorisation of such evacuation-relevant data. The review of the techniques is carried out to investigate how to develop methods for each stage of the pipeline. These stages are illustrated in Figure \ref{fig:pipeline} and explained below. In addition to the explanation, we outline the focus of the literature review for each stage. 

Figure~\ref{fig:pipeline} presents a structured overview of the data mining pipeline, segmented into three key stages: \textbf{Data Collection}, \textbf{Data Cleaning}, and \textbf{Data Categorization}. Each stage reflects a critical phase in transforming raw social media content into actionable insights, with methodological choices presenting distinct trade-offs between precision, recall, and computational efficiency \cite{Bruns2012,Paul2013,Zhou20223213}, summarized as follows:

\begin{itemize}
    \item \textbf{Data Collection}: This initial phase involves extracting posts from online conversations using hashtags and keywords. The iterative nature of this process---where extracted posts help identify additional hashtags and keywords---aligns with common information retrieval techniques and bootstrapping methods discussed in the review.

    \item \textbf{Data Cleaning}: This stage addresses the challenge of noisy and ambiguous data. Feature extraction and active learning are employed to train classifiers that can distinguish relevant content. When classifiers are uncertain, human judgment is incorporated to refine the dataset. This hybrid approach reflects semi-supervised learning strategies and human-in-the-loop systems highlighted in the reviewed literature.

    \item \textbf{Data Categorization}: In the final stage, the cleaned data is categorized using further feature extraction and active learning. The categorization answers specific questions---such as ``which bushfire'' and ``where''---which mirrors the use of structured information extraction and event detection techniques covered in the review.
\end{itemize}

A typical process of collecting social media data has the following steps \cite{Efron2010787,Bozarth2022}. It uses hashtags and keywords to initiate a search for data, but in later steps, it also employs a geo-location search. The hashtags and keywords may lead to some geotags, which are coordinates of locations. A bushfire affects nearby people, as well as people in the location of a geo-tag. The places through which the evacuees pass are considered nearby places. Thus, each discovered geotag can be integrated into a search within a neighbourhood of its location to fetch posts in the neighbourhood. As shown in the first stage of Figure \ref{fig:pipeline}, the process iteratively expands its selection of hashtags and keywords by extracting hashtags / keywords from the sample data automatically. It iterates for several rounds until the collected messages are sufficiently complete with respect to their coverage of the selected hazard and evacuation. The focus of this scoping review in data collection is on strategies to select hashtags and keywords that are relevant to evacuations during natural hazards. 

While data collection in the pipeline ensures high information retrieval recall, data cleaning ensures good precision. To clean the data, supervised classifiers are trained so that noisy social media data can be classified into two parts: on-topic and off-topic parts  \cite{Calisir2018115}. Only the on-topic part of the data can be forwarded in the pipeline for further data categorisation. The on-topic data are those mentioning the occurrence of a bushfire or an evacuation from a bushfire. 
An adaptive sampling technique of active learning (e.g. \cite{Lookman2019,Torres2019,Bhosle202072,Batista2022,El-Hasnony2022}) can reduce the cost of data labelling for training supervised classifiers.
As illustrated in the second stage of Figure \ref{fig:pipeline}, to apply the technique and identify the on-topic data, the characteristics of the bushfire evacuations must be extracted as features from the data. Thus, the focus of the literature review on data cleaning is on the extraction of these features.

The categorisation of the data in the pipeline of Figure \ref{fig:pipeline} categorises the data by six dimensions: bushfire event, evacuee, behaviour, time, location, and mode of transportation. First, it regroups the social media messages according to the social media events \cite{Zhou2014381,Zhou20223213} which correspond to different bushfire incidents. Then, it divides them according to different kinds of user behaviours, such as deciding to leave or stay, preparing to evacuate, moving to safety, taking shelter or refuge during a bushfire (destination choice), and returning to the area (if possible) after the bushfire. Thus, these behaviours include those of the four-step model mentioned in Section \ref{sec:overview}. Other dimensions are attached to specific behaviours, so the behaviours determine which step of the evacuation model to which they correspond. These dimensions can reveal detailed information of behaviours when clustering the data according to other dimensions. If a message contains the time and location of the behaviour, direct extraction of time and location can be performed. Otherwise, the time and location can be estimated from the message posting time (acknowledging that this is not necessarily the behaviour time), the conversation context, and the bushfire event. In particular, departure time, destination choice, and route choice are extracted from these two dimensions. The last dimension, the transportation mode can include travel by foot, public transport (e.g., bus or train), personal vehicle, or possibly other types of transportation. The categorisation can also adopt active learning for interactive accuracy improvement. Active learning is a kind of machine learning that allows the learner model to select the most informative data points for labelling, rather than using a random or predefined subset of data. Active learning can reduce the amount of labelled data needed to train an accurate model, which can lower the cost and time associated with data labelling. The accuracy of categorisation can affect the effectiveness of identifying behaviours. The focus of the literature review on data categorisation is on social media event detection, evacuee identification, evacuation behaviour classification, time and location estimation, and identification of departure time, destination choice, route choice, and transportation mode.

The selection of techniques across all pipeline stages involves navigating key trade-offs. \textit{Traditional methods} (e.g., keyword-based filters, rule-based systems, simpler statistical classifiers) are often more interpretable, computationally efficient, and require less training data. However, they may lack the flexibility and power to handle the noise, volume, and unstructured nature of social media data. \textit{Machine learning (ML) and deep learning approaches} excel at capturing complex patterns from large datasets and can adapt to new slang or emerging events. Yet, they often act as "black boxes," require large amounts of labeled data, and are computationally intensive and prone to overfitting on specific events. The choice between these paradigms is not mutually exclusive and often depends on the specific task, available resources, and the desired balance between interpretability and predictive performance. Subsequent subsections will detail these trade-offs within each stage of the pipeline.

\subsection{Data Collection \label{ssec:data_collection}}
In this subsection, we review the literature on the collection of social media data for the examination of bushfire evacuation. The aim is to understand how to select hashtags and keywords to find suitable data. 

Social media platforms usually offer Application Programming Interfaces (APIs) to allow users to collect data, and the proposed data collection module can adopt those APIs. There are often {\bf three types of APIs}, and typical social media platforms provide all \cite{Pfeffer2018}. The first type of API is {\bf search API}, which allows a user to find events using hashtags, keywords, time ranges, locations, etc. The search API is interactive. A user can examine the search results of some keywords that they use in the search API. If the user is not satisfied with the search results, they can change the keywords and search again. The second type of API is {\bf REST API} (i.e., Representation State Transfer API), which offers CRUD (Create, Retrieve, Update, Delete) operations to the data on a social media platform. REST API enables a computer programme developed by the user to automate the activities of the user account, such as reading and analysing posts and even writing posts to talk to other users. The third type of API is the {\bf streaming API}, which continuously pushes live messages to the user programme if the programme connects to the streaming API endpoint. The connection can be customised by selecting the conversations or the timelines to listen to. The conversations or timelines are specified with hashtags, keywords, and/or locations. However, it is difficult to know about an emerging event until the event has received a great deal of attention. A general solution to this issue is a combined use of three APIs. First, conversations regarding the event are pinpointed via the search API. Then, past posts in the conversations are read and analysed using the REST API. Finally, conversations are connected using the streaming API to continue receiving future posts. 

\begin{figure*}[htb!]
    \centering
    \includegraphics[width=0.7\linewidth]{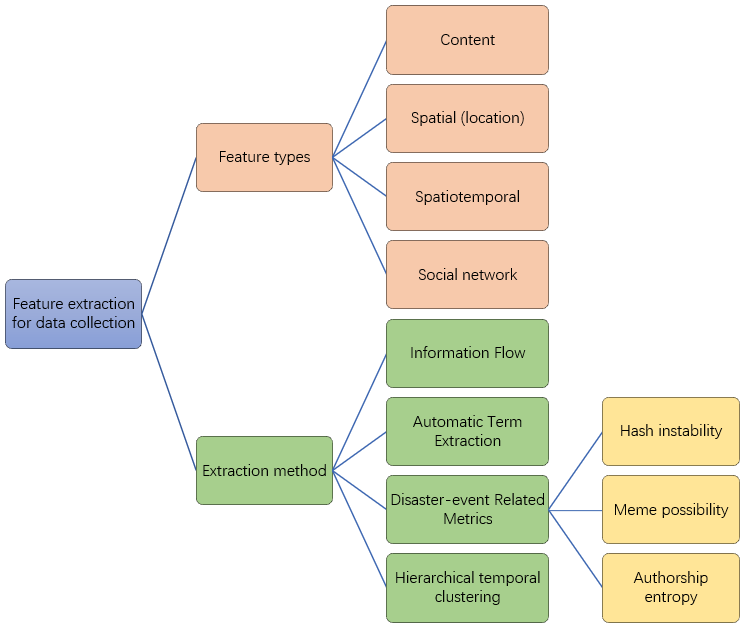}
    \caption{Feature Extraction for Collecting Data in Social Media Data Mining of Human Behaviours during Bushfire Evacuations.}\vspace{-2ex}
    \label{fig:data_collection}
\end{figure*}

Since full-archive data of a social media platform are usually prohibitively large in volume, the event pinpointing needs search strategies, many of which have been studied in existing works. These works extract features for data collection. As shown in Figure \ref{fig:data_collection}, they can be divided into two groups: one group focusses on feature types, the other group focusses on the extraction method. 

The feature types for data collection include content-based (i.e., texts), spatial-based (i.e., locations), spatiotemporal-based (i.e., dynamic movements with time-varying locations) and social network-based (i.e., social relations) features.
\textcolor{black}{Bruns et al 2012} \cite{Bruns2012} suggested that new data should be captured from a social media platform to adapt the {\bf search content features} such as hashtags and keywords via analysis of the feature distribution, the rise and fall of a feature's occurrence over time, and the co-occurrences of multiple features. \textcolor{black}{Paul et al 2013} \cite{Paul2013} demonstrated that sampling location-specific data can help to quickly find hazard-specific priority areas, which means that a search strategy should use {\bf location features} in addition to content features in the search. \textcolor{black}{Zhou et al 2022} \cite{Zhou20223213} proposed yet another adaptive metric to extract not only content features, but also {\bf spatiotemporal features} and {\bf social network features} that are critical in the early discovery of hazard events.  Generally speaking, content features and social network features are good at discovering bushfire-relevant data on social networks, but their coverage of relevant messages for events is often limited. On the other hand, spatial and spatio-temporal features ensure better coverage, but they are limited in the discovery of relevant data.  

To extract features, \textcolor{black}{Marcus et al 2011} \cite{Marcus20111259} developed a method to detect hazard events on a social media platform using spikes of {\bf information flow}. More importantly, its method adopts {\bf automatic feature extraction} to label events (e.g. using the term frequency invert document frequency metric \cite{Baker1962512} for extraction). This method can be used to monitor the information flows and automatically extract features. \textcolor{black}{Joseph et al 2014} \cite{Joseph2014672} measured the effectiveness {\bf in detecting hazard events} of various features and found that a weighted overlap metric has a strong correlation with effectiveness. The weighted overlap metric measures the overlap of a feature inside and outside the event. According to the conclusions of \textcolor{black}{Joseph et al 2014} \cite{Joseph2014672}, automatic feature extraction should not be limited to their algorithm. It can also adopt more {\bf hazard event-related metrics} and even {\bf adaptive metrics} based on machine learning models. \textcolor{black}{Cui et al 2012} \cite{Cui20121794} proposed three other effective metrics for detecting broken events, that is, hashtag instability, meme possibility, and authorship entropy. {\bf The hashtag instability} is a quantified metric to detect spikes in information flows, which is significantly higher in hazard events than in daily events. {\bf The possibility of meme} is very helpful in distinguishing hazard events from meme events, which are popular (i.e., they have similar spikes) thus working similarly to any other social media trends. {\bf The authorship entropy} enables to distinguish another type of spiky event, that is, spam. A hazard event usually has a high authorship entropy, whereas spam has a low authorship entropy. Events are constantly evolving and migrating \cite{Wang201710,Zhou20223213} so it is important to use adaptive metrics.
\textcolor{black}{Xing et al 2016} \cite{Xing20162666} developed an adaptive metric and an algorithm, {\bf mutually Generative Latent Dirichlet Allocation}, which can efficiently extract content characteristics. \textcolor{black}{Wang et al 2017} \cite{Wang201710} designed a similar adaptive metric based on the profile and similarity of hashtags, which can support {\bf hierarchical temporal clustering} of social media streams and automatically extract important hashtags. \textcolor{black}{Martín et al 2016} \cite{Martín2016} and \textcolor{black}{Erz et al 2018} \cite{Erz201848} both pointed out that social posts with hashtags are likely to attract more followers than posts without hashtags. Thus, hashtags should play a central role in content features, in particular when detecting {\bf emerging events} (i.e. events that happened recently and have not been noticed by many people).

We review the literature on the features of social media messages for data cleaning. These features are characteristics that are used to classify whether a message is relevant to a bushfire or an evacuation from the bushfire. Note that data collection is also based on features, but the features for data collection and the features for data cleaning have different purposes. The features for data collection describe sufficient conditions to ensure high data retrieval recall. The features for data cleaning specify the necessary conditions to guarantee high precision. To specify the necessary conditions, the features for data cleaning should increase the awareness of the situation of human annotators for bushfires and bushfire evacuations. Situation awareness is the ability to sense the situation, which is a bushfire or an evacuation from a bushfire in our case. Increasing situation awareness makes this detection easier. Features that allow for high situation awareness can benefit human annotators, as well as classification algorithms. Posts are short and informal, so judging whether a post is relevant to a bushfire is often difficult based only on the post. If some highly relevant features of bushfires or bushfire evacuations are displayed on the annotation user interface to increase situation awareness, human annotators can have more information and make judgement easier. However, too many features on the user interface can overload the information and hinder the efficiency of judgment. Thus, the features must be carefully chosen. The situation is similar to the classification algorithms. If a human can make judgment easier, a classification algorithm can also be simpler. 

\paragraph{Critical Analysis $\&$ Trade-offs:} The primary challenge in data collection is balancing {\bf recall} (retrieving all relevant posts) against {\bf precision} (avoiding irrelevant noise), while also managing {\bf computational cost} and {\bf adaptability} to the evolving language of a crisis. 

This challenge manifests in a clear trade-off between \textit{traditional} rule-based methods and \textit{machine learning (ML)}-driven adaptive techniques. \textit{Traditional methods} (e.g., static lists of keywords/hashtags, fixed geographic queries) are computationally inexpensive, simple to implement, and offer full transparency. However, they are brittle; they fail to adapt to novel terminology or unforeseen event dynamics, leading to rapidly decaying recall and precision. Conversely, \textit{adaptive ML techniques} (e.g., iterative feature expansion, topic modeling) are designed for dynamic environments. They automatically discover emerging relevant content, maintaining high recall over time. This advantage comes at the cost of significant computational resources, implementation complexity, and reduced interpretability.

\textbf{Which method works best under which conditions?}
For \textbf{rapid, retrospective analysis} of a specific, well-defined event, a carefully crafted \textit{traditional} search strategy is often sufficient and most efficient. For \textbf{building real-time monitoring systems} or studying prolonged, complex events where public discourse evolves, the investment in \textit{adaptive ML-driven collection} is necessary and justified to ensure comprehensive and relevant data coverage.

\subsection{Data Cleaning \label{ssec:data_cleaning}}
\begin{figure*}[htb!]
    \centering
    \includegraphics[width=0.7\linewidth]{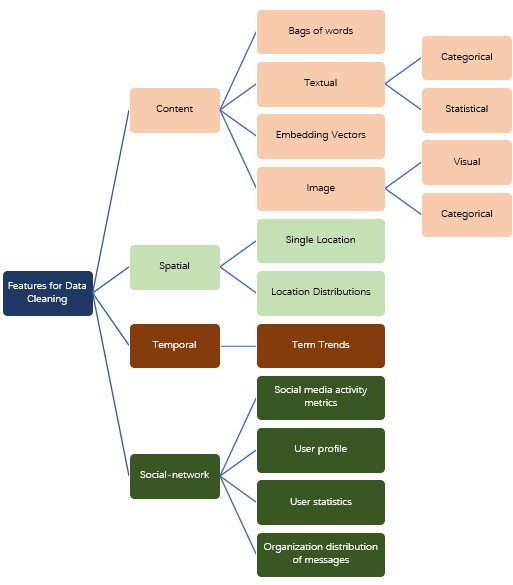}
    \caption{Features for Cleaning Data in Social Media Data Mining of Human Behaviours during Bushfire Evacuations.}\vspace{-2ex}
    \label{fig:data_cleaning}
\end{figure*}

As shown in Figure \ref{fig:data_cleaning}, there are four types of features related to situation awareness: content features, spatial features, temporal features, and social-network features. Content features are derived from the text of each social media message. The spatial features are about the locations where these messages are posted. Temporal features are about trends. The social network features are about how users maintain a social network.

Most works we reviewed for this subsection are about content features, as content is the easiest standard to judge whether a message is relevant to a bushfire or an evacuation from a bushfire. Work about content features can be divided further into several groups, including bags of words, textual categorical features, 
textual statistical features,
embedding vectors,
image visual features, and image categorical features.
Bags of words \cite{Cameron2012695,Cobo20151189} are used in natural language processing and information retrieval to represent text with the distribution of words, as the distribution of words is often a good representation of semantic context. 
Textual categorical features include hashtag or keyword categorical features \cite{Abel2012305,Chowdhury2020498,Döhling2011127} and
topics \cite{Balech2021308,Imran20161638,Kanth201982,Ashktorab2014354,Dereli2021,Kemavuthanon2020212}, where topics are usually obtained from clustering or classification of semantic contexts.
Textual statistical features include statistical scores \cite{Olteanu2014376}, stability metrics \cite{Chowdhury20151227}, and mutual information \cite{Hodas20151201} of texts.
Embedding vectors \cite{Bandyopadhyay2018925,Dasgupta201671} are vectorised representations of semantic context. Different from bags of words that often need a very large number of dimensions of a vector space to represent a context due to a large number of words and phrases occurring in a corpus, embedding vectors are compressed representations that only need a relatively small number of dimensions of the vector space. Bags of words, textual features, and embedding vectors are all derived from the text of a message. The image features are derived from the image or video of a message. Visual features of the image \cite{Weber2020331} are statistical representations of the characteristics of the image. Image categorical features  \cite{Jony2019,Ilyas2014417} are obtained from the clustering or classification of images.

A small number of works reviewed in this subsection adopt other types of features.
Work about spatial features can be divided into locations \cite{Abel2012305} and spatial distribution of posts  \cite{Hernandez-Suarez2019}. A work using temporal features \cite{Balech2021308} uses term trends. 
The work on social network features includes social network activity metrics \cite{Bruns201391}, user profile \cite{Abel2012305}, user statistical features \cite{Cobo20151189}, and organisational distribution of posts \cite{Opdyke201486}. These are auxiliary features that can be combined with content for better data cleaning.

Generally, content features provide more semantic information than other features, such as a user's emotions, opinions, decisions, and actions. However, content feature extraction is often computationally expensive and in turn requires more effort from human annotators to be understood. Spatial and temporal features offer straightforward information about location and time, but spatial features can be sparse in social media. Also, temporal features are noisy because the time associated with a social post is often different from the time of the behavioural occurrence described in the message. Social media features enable a deeper understanding of the social relations of people involved in an event, such as their roles in a bushfire evacuation, but the relations can be dynamic, temporary and complicated.

\paragraph{Critical Analysis $\&$ Trade-offs:} The core trade-off in data cleaning is between the high accuracy and automation potential of \textit{supervised machine learning classifiers} and the lower resource requirements of \textit{traditional} manual or rule-based filtering.
\textit{Traditional} approaches (e.g., manual coding, keyword blacklists/whitelists) are transparent and require no training data, making them suitable for small datasets or initial exploration. Their performance plateaus quickly and they do not scale. \textit{Supervised ML models} (e.g., SVM, neural networks) can automate cleaning at scale with high precision once trained. Their major drawback is the dependency on large, accurately labeled datasets, which are expensive and time-consuming to create.

\textbf{Which method works best under which conditions?}
The choice is often dictated by data volume and resource availability. \textbf{Small-scale studies} or projects in the early \textbf{prototyping phase} may rely on \textit{traditional} manual methods. \textbf{Large-scale analyses} necessitate investing in \textit{supervised ML}. A hybrid strategy is often optimal: using \textit{active learning}—an ML technique that prioritizes data points for human annotation—to minimize labeling effort while maximizing classifier performance, thus effectively addressing the data challenge of annotation cost.

\subsection{Data Categorization \label{ssec:data_categorization}}
\begin{figure*}[htb!]
    \centering
    \includegraphics[width=0.55\linewidth]{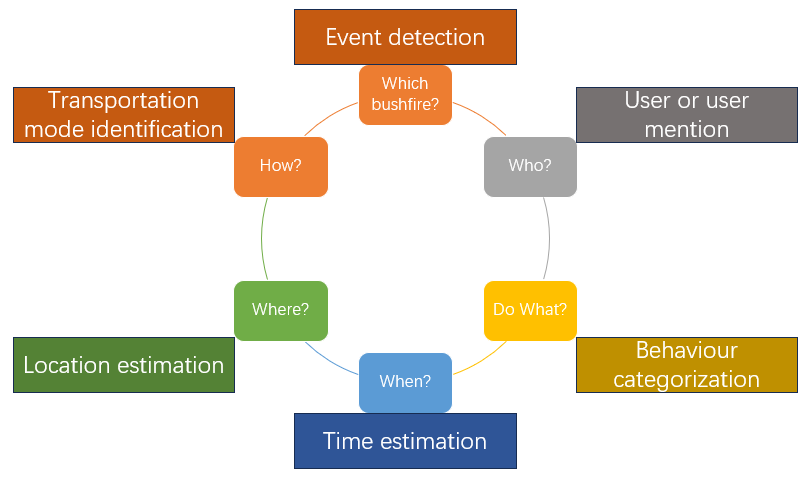}
    \caption{Features for Cleaning Data in Social Media Data Mining of Human Behaviours during Bushfire Evacuations.}
    \label{fig:data_categorization}
\end{figure*}
In this subsection, we review the existing works for categorising social media messages along six dimensions: 
\begin{itemize}
    \item {\bf which bushfire} can we study, 
    \item {\bf who} are involved in the evacuation, 
    \item {\bf what} do they do during the bushfire, 
    \item {\bf when} do they do it, 
    \item {\bf where} do they go (e.g., for safety) 
    \item {\bf how} do they get there.

\end{itemize}
The dimension of ``which bushfire'' is extracted by event detection on social media. Given a set of social media messages to an event detection method, the method identifies some events and finds a subset of messages for each identified event. In our scenario, the method finds a subset of relevant messages for each identified bushfire event. Therefore, if a message is relevant to a bushfire, its category in the ``which bushfire'' dimension can be the relevant bushfire; otherwise, if a message is irrelevant to any bushfire, its category in the ``which bushfire'' dimension can be null.  The category in the dimension of ``who'' is extracted either from the user who posted this message or a person mentioned in this message. This is a relatively simple data extraction task that can be handled using classifiers such as Support Vector Machines (SVM\cite{Cortes1995273}). The category in the dimension of ``do what'' is about the behaviour described by the message, which can be extracted from texts from social media posts using classifiers. The cost of annotating training data for these classifiers can be reduced by three other types of behaviour categorisation methods: spatiotemporal analysis, sentiment analysis, and social network analysis. The fourth and fifth dimensions, i.e., ``when'' and ``where'', are about time and location. Although each message has a time stamp, the time stamp is often after the behaviour described in the message, so time estimation is required. Location information in a message is often missing, so location estimation is also required. The last dimension, i.e., ``how'', refers to the mode of transportation of the evacuation, which can be extracted from text or deducted from conversation context. The other ``how'' of an evacuation timeline, i.e., the choice of evacuation route, can be inferred from the time and location information of a sequence of messages, so it is not categorised separately.

The bushfire event is the primary dimension of the categorisation of the data. Tracking the development of bushfire events is based on this dimension. The categorisation along this dimension can be based on a recent event detection method of \textcolor{black}{Zhou et al 2022} \cite{Zhou20223213}, which constructs a maximal user influence graph from the social-user relationship to identify event migrations. 
Other existing event detection methods can be grouped into non-location-constrained \cite{Cao20213383,Peng2021,Xing20162666} and location-constrained approaches 
\cite{Chen2018516,Kim2011529,Sakaki2010851,Singh20101181,Yin2013661,Yin201252,Zhou2014381}. 
Non-location constrained methods, such as the Pairwise Popularity Graph Convolutional Network model \cite{Peng2021} and Knowledge-Preserving Incremental Heterogeneous Graph Neural Network (KPGNN \cite{Cao20213383}), lack handling of location information, thus are inapplicable to space-sensitive social events. 
Thus, location-constrained methods have been proposed to incorporate location as an attribute of social media for space-sensitive applications. Examples include probabilistic spatio-temporal models for earthquake event location \cite{Sakaki2010851}, user interest levels at different geolocations such as social pixels \cite{Singh20101181}, visual tools for understanding topic movement over time and space \cite{Kim2011529}, and user reposting behaviour for event evolution \cite{Chen2018516}. 
These techniques fix time or location in a spatio-temporal range, making them infeasible for non-consecutive event migration scenarios. Studies have also been conducted to detect patterns in geotagged posts for social events \cite{Chen2016270,Choi2018,Huang2018}, but these methods only handle simple user behaviour movements within short time intervals. 
Topic-model-based visual analytics systems like TopicOnTiles in \cite{Choi2018,Huang2018} focus on small-scale spatial-temporal events and their textual content, but they cannot handle complex events with non-consecutive migrations over time and space. Unlike those methods, the event detection method proposed in \cite{Zhou20223213} can handle complex events with migrations. Since bushfire events are complex events, the techniques and ideas in \cite{Zhou20223213} might not be useful for our motivation. 

Evacuation behaviours are, first of all, about leaving or staying. For evacuees, departure time, destination choice, and transportation mode are three important decisions. Those staying must decide whether to defend their property or find a safer local shelter. Behaviours can always be classified using the text of a social media message. Text-based classifiers must be trained with training data \cite{csahin2019emergency}. We found three types of analysis that might reduce the cost of annotating these training data: spatiotemporal analysis, sentiment analysis, and social network analysis. Thus, we may still need to manually annotate training data for the classifiers, but these three types of analysis can automate part of the annotation. 

The spatiotemporal analysis for evacuation behaviour classification is based on GPS data. Discrete GPS locations such as geotags on microblogging services are useful since big data containing many discrete GPS locations can be used to estimate routes and speeds. Social media platforms such as microblogging services provide discrete GPS locations in the form of geo-tags and location hashtags, the latter of which need conversion to turn into discrete GPS data. \cite{Li2017,Wu2020,Kumar20181613}  provided detailed methods to analyse GPS data for understanding human behaviour.

Sentiment analysis for evacuation behaviour classification can find emotional factors affecting the sequence of decisions made during an evacuation.  \cite{Berger2012,Kramer2014,Lu2021} explained the cause of collective emotional fluctuations during hazards and evacuations. \cite{Berger2012,Kramer2014,Yabe2019} demonstrated methods to analyse the impact of these sentiments on evacuation behaviours. \cite{Moussaïd20155631} provided quantitative insights into the public response to risk and the formation of often unnecessary fears and anxieties that are amplified during information propagation. Many factors are public behaviours not directly related to an evacuation, but can indirectly affect the evacuation. \cite{Ukkusuri2014110} proposed a method to find these public behaviours, also using sentiment analysis. Algorithms that allow sentiment analysis on social media data include \cite{Thelwall2012163,Buscaldi20151185,Bai20161177,Beigi2016313,Alfarrarjeh2017193,Ragini201813,Hassan2019104,Mendon20211145,Hassan2022}. In particular, \cite{Thelwall2012163} focusses on detecting the strength of sentiment in social media content. It explores methods to measure the intensity of sentiment expressed in online text, particularly on social media platforms. \cite{Buscaldi20151185} investigates sentiment analysis techniques applied to microblogs during natural hazards, using the 2014 Genoa floods as a case study. The authors explore how sentiment analysis can contribute to hazard management. \cite{Bai20161177} proposes an approach using Weibo text analysis to monitor post-hazard incidents. It focusses on the negative sentiment analysis in Weibo messages to gain insight into hazard situations. \cite{Beigi2016313} provides an overview of sentiment analysis in social media and its relevance to hazard relief efforts. It discusses how sentiment analysis can aid in understanding public sentiment during hazards and aid relief operations. \cite{Alfarrarjeh2017193} focusses on geo-spatial multimedia sentiment analysis during hazards. The authors explore how sentiment analysis can be applied to multimedia data (images, videos, etc.) to gain insights into public sentiment and emergency response. \cite{Ragini201813} discusses using big data analytics and sentiment analysis for hazard response and recovery efforts. It emphasises the role of sentiment analysis in managing and addressing hazard-related information. \cite{Hassan2019104} focusses on sentiment analysis applied to images of natural hazards. The authors explore techniques to extract sentiment information from images, improving the understanding of public emotions and reactions during hazards. \cite{Mendon20211145} proposes a hybrid method that combines machine learning techniques with lexicons for sentiment analysis of microblogging services data related to natural hazards. The study focusses on analysing microblogging services data specifically related to natural hazards. The hybrid method aims to improve the understanding of emotions and sentiments expressed during such events. \cite{Hassan2022}  introduces an approach to automatically determine the sentiment associated with images captured during hazards. This involves detecting emotions and sentiments expressed in images to provide insights into the public's emotional responses to hazards.

The social network analysis for the classification of evacuation behaviour reveals social factors that affect evacuation decisions. Such social factors arise from communication with connected peers on social networks. \cite{Starbird2010,Abdullah2017432} point out that reposts highlight social factors on social media by repeatedly showing microblogging services users the content of the original posts. \cite{Starbird2010} provided an algorithm to track reposts. \cite{Li201834} classifies the messages that contain these factors into five types. The method in \textcolor{black}{Li et al 2018} could be useful to track the source of the information on the factors. \cite{Jiang2019} proposed a method to find the participation of evacuees in social network communities.  \cite{Lu2014} suggested social network communities define collective human behaviours that can ultimately affect behaviour during emergencies. Many social network analyses require network community detection algorithms such as \cite{Blondel2008,Que201528,Zeng2018268,Ghosh2019,Bhowmick202043,Samir2021,Sattar202210275,Gao2023}. In particular, \cite{Blondel2008} presents a "fast unfolding of communities" algorithm to identify communities within large networks. The algorithm is based on modularity optimisation and aims to efficiently detect communities in networks. \cite{Que201528} introduces the "Louvain Algorithm" for scalable community detection. The algorithm focusses on optimising modularity and is designed to efficiently handle large networks. It uses a greedy approach to iteratively merge or split communities. \cite{Zeng2018268} presents a distributed version of the Louvain algorithm for community detection in large-scale graphs. The focus is on scalability, and the algorithm is designed to run on distributed computing environments to handle very large networks. \cite{Ghosh2019} investigates the scaling and quality of different modularity optimisation methods for graph clustering, and compares the performance of various methods in terms of their ability to handle large graphs while maintaining cluster quality. \cite{Bhowmick202043}  introduces "LouvainNE," a hierarchical Louvain method for network embedding. The method aims to create high-quality and scalable solutions for various applications. \cite{Samir2021} presents an approach called "LKG" for the problem of influence maximisation in social networks. It is community-based and focusses on efficiently identifying influential nodes in large social networks. \cite{Sattar202210275} introduces a distributed scalable version of the Louvain algorithm for the detection of communities in large graphs. The algorithm is designed to be efficient in handling large datasets while maintaining accuracy. \cite{Gao2023} presents an algorithm for "Automatic Detection of Multilevel Communities" that is scalable, selective, and resolution limit-free. The algorithm identifies communities at multiple levels of granularity in networks.

Each of the three types of spatio-temporal analysis has its own pros and cons. Spatial-temporal analysis is good at extracting evacuation behaviour but not as helpful in analysing the factors affecting the behaviours. The other two types are not as helpful in identifying evacuation behaviours, but sentiment analysis can reveal emotional factors of the behaviours and social network analysis is better at revealing social factors that affect the behaviours. Fortunately, \cite{Shekhar20151719,Chen2019436,Yabe2021} demonstrate that these analyses can work together.

Social media messages generally have time stamps, but the time stamps are often after evacuation behaviours have been performed because people may not report their behaviours as they are happening. Furthermore, only a small fraction of social media posts have geotags due to the protection of user privacy. However, information on time and location is required for this type of work. Thus, they are estimated whenever they are missing. In other words, if the departure time, destination, and route choices are found in social media data posts, we extract them from the posts. Otherwise, we try to estimate them according to the conversation context of the posts. Time estimates and movement trend prediction are closely connected. The movement trend is the relation of location and time, since from the past sequence of location and time, we can expect the location in a future time.  \cite{Paterson201925} proposed a probabilistic method that produces direct time estimates from movement trends. \cite{Oga2020109} developed a regression-based method that provides trend prediction.  Most location estimate methods are content-based. The only social network-based method for location estimation is \cite{Yamaguchi2013223}. Many content-based methods need more than social media data to estimate location. For example, \cite{De_Longueville200973,Gelernter2011753} need a gazetteer (i.e., a geographical dictionary or directory used in conjunction with a map or atlas), \cite{Ikawa2012687} needs external location services such as Foursquare and LocTouch, \cite{Lingad20131017} needs Yahoo Placemaker (a freely available geoparsing Web service that converts the name of a place to a pair of geocoordinates), \cite{Krishnamurthy2015187} needs Wikipedia, and \cite{Salfinger2016212} needs geonames.org (a website offering a geographical database covers all countries and contains over eleven million place names that are available for download free of charge). These external dependencies might incur runtime overhead due to external data fetching, unless the external data are already stored locally. The runtime overhead of external dependencies can lead to high cost in the big data analysis of behaviours if it is not handled properly.

During evacuation movements, people have various choices in transportation mode, such as walking, private vehicles, public road transport, and other transport options (via sea or air). In some social media messages posted by evacuees, transportation modes are described directly in the message. For these messages, the transport modes can be easily identified and classified by a content classifier. Few existing works \cite{Maghrebi20161648,Raczycki2021} on the classification of these messages were found during our literature review. \cite{Maghrebi20161648} classify transport modes using keyword matching. \cite{Raczycki2021} uses regular expression matching to classify transport modes. In the other messages, transportation modes are missing, but can be inferred to some extent. In particular, if an evacuee complained about traffic congestion or a traffic incident on a social media post, they are likely travelling by private vehicle or public road transport. Thus, we consider studies on traffic congestion or incident detection in \cite{Sakaki2012221,D'Andrea20152269,Jones2018} to be related to the classification of transportation modes. \cite{Sakaki2012221,D'Andrea20152269} both adopt support vector machine in such a classification, though \cite{Jones2018} showed the existence of a better classifier.  

\paragraph{Critical Analysis $\&$ Trade-offs:} Categorization tasks are dominated by a trade-off between the flexibility and power of \textit{ML} and the controllability and reliability of \textit{traditional} methods.
\textit{Traditional methods} like rule-based classifiers (e.g., using regular expressions to find transportation modes \cite{Raczycki2021}) or dictionary-based sentiment analysis are interpretable and work reliably on specific, predictable patterns. They are not designed for the nuanced, implicit, or novel expressions common in social media. \textit{Machine learning approaches}, from simpler models like SVM to advanced deep learning and LLMs, capture complex linguistic patterns and generalize much better to unseen data. The trade-off is their black-box nature, massive data requirements, and computational cost.

\textbf{Which method works best under which conditions?}
Use \textit{traditional rule-based methods} for well-defined, concrete categories where precision is paramount and expression variety is low (e.g., detecting explicit mentions of vehicle types). Use \textit{ML} for nuanced, subjective, or complex categorization tasks (e.g., inferring emotional state, identifying protective actions from descriptive text, event detection from context). For tasks like location estimation, \textit{traditional} gazetteer-based methods may be more accurate for named places, but \textit{ML} methods are essential for interpreting imprecise or relative locations (e.g., "near the big fire on the hill").

\section{Applications of social media \label{sec:applications}}
In the Introduction, we briefly mention possible applications of social media mining of human behaviours during bushfire evacuations. This section examines four key applications in detail. The literature reviewed related to applications consists of papers on past applications for any evacuation (not just bushfire evacuations) using data collected by traditional means, such as quantitative surveys and manual observations. It also includes speculations on possible additional uses of social media that have not been explicitly identified yet in the literature.

\subsection{Evacuation Model Calibration and Validation}

This work showed that social media data can contain useful information related to several phases of an evacuation, i.e. from the evacuation decisions itself to the evacuation movement process. In this context, evacuation models are useful tools that can allow predicting evacuation responses and times in a given event. The accuracy and reliability of such models will be as good as the underlying models adopted for their calibration and validation as well as the intrinsic modelling assumptions used for prediction. The use of social media data for evacuation modelling can be dual. On one hand, it can aid the calibration of evacuation modelling inputs, by providing useful information on all the modelling layers and sub-layers implemented by a model (e.g. each of the four-step process of an evacuation model).  On the other hand, it can be used to perform validation studies of evacuation simulations, i.e. evaluate if an evacuation model prediction is an accurate representation of a real world scenario \cite{ronchi2021wui,Ronchi20231493}. This would require validation testing of an evacuation model with fire evacuation scenarios which are representative of the conditions being represented in the scenario of interest. 

To date, calibration and validation can be conducted using a range of data such as survey/questionnaire data \cite{Toledo20181366,Katzilieris2022}, community evacuation drill data  \cite{Gwynne2023879}, traffic detector data  \cite{Rohaert2023,Rohaert2023a}, connected vehicle data  \cite{ahmad2023evaluating}, or GPS data \cite{Zhao2022,Wu2022b}. Social media data are potentially extremely useful to complement such data-sets for calibration and validation efforts. This can relate to all modelling layers that are generally included in a coupled WUI fire evacuation model along with how they are influenced by the fire evolution. Evacuation simulation tools make use indeed of different modelling approaches, including sophisticated behavioural models for the representation of evacuation decisions and actions during hazard evacuations. Depending upon the modelling method; i.e., macroscopic, microscopic,  mesoscopic, the models require different types of data at different temporal and spatial scales of the event \cite{ronchi2017sanctuary}.  Evacuation models can typically represent one or more modelling layers which affect evacuation (human response, pedestrian movement, evacuation via a transport mode, impact of wildfire spread on evacuation \cite{Ronchi2019868}). Examples of such models include WUI-NITY \cite{Wahlqvist2021} and SEEKER \cite{Singh202273}. 

Social media data could for instance be used to test the key modelling assumptions adopted by a model in its representation of the stay vs evacuate decision and departure time modelling (e.g. trip generation modelling). Similarly, movement modelling (both for people walking on foot, via private vehicles or any other transportation mean) could be benchmarked with social media data providing information on the trip distribution, modal split and traffic assignment stages.  Given the overall scarcity of datasets available for all bushfire evacuation decisions \cite{Kuligowski2021}, social media data can provide valuable insights that can lead towards more accurate predictive capabilities of evacuation modelling tools. This can in turn provide enhanced decision support to all parties involved in bushfire evacuation planning. 

\subsection{Emergency Communication \label{ssec:planning_application}}
An application for social media data on human behaviour during fires is the design of emergency communication for communities. Community-scale emergency communication can include information on the location of identified evacuation shelters (and other safe destinations),  routes for timely evacuation, and mitigation strategies (e.g., traffic management solutions or places of last resort). Social media data can also be used to identify the most appropriate means for emergency communication and investigate manners to increase instruction compliance (e.g. to mandatory evacuation orders).  Social media data can help understanding the effectiveness of different emergency communication strategies and in turn provide help in designing them.  If informed by the expected human behaviour, emergency communication can in fact lead to increased safety by comprehensively consider a credible set of 'what if' bushfire and resulting evacuation scenarios. A better understanding of human behaviour during the evacuation of the community in question is a necessary step towards the development of effective emergency communication strategies  \cite{Doermann2021815,Kuligowski2023}. This can greatly benefit from the analysis of previous responses and compliance to instructions provided by authorities.

\subsection{Personalised Evacuation Training \label{ssec:education_application}}
Evacuation training programmes are designed to educate individuals or communities about the necessary steps and strategies for safe evacuation in the event of a bushfire. Bushfires can pose significant threats to human lives, animals and property, especially in areas with dense vegetation and dry conditions. A possible application for the use of social media to better understand evacuation behaviour is to inform the development of evacuation training \cite{Liu2023,Gao2022,Lin2020,Moussaid2016,Kinateder2016}. Social media data mining of human behaviour during bushfire evacuations supports this application by helping understanding the population under consideration. This includes estimating the level of technology literacy (which is important for innovative training applications such as the use of Virtual Reality discussed in \cite{Menzemer2023}) and evaluate behavioural factors which may have an impact on training effectiveness. This will help ensuring that the unique circumstances of each individual or community are taken into account when developing or using a given training approach. This is likely increasing the likelihood of successful evacuation. In other words, social media data can help identifying typical user profiles in a given community, and linked it with their expected behaviours. This information can be used for the development of  effective and personalised fire safety training programmes. Tailored fire safety training programme could be focused on educating people living in bushfire-prone areas on several aspects related to the evacuation process (i.e. evacuation decision, route/destination choice, interaction with authorities)

\subsection{Resource Allocation for Evacuation Preparedness \label{ssec:resource_application}}
Resource allocation for bushfire evacuation preparedness involve the planning, coordination, and allocation of various means to ensure a safe and efficient evacuation process during a bushfire emergency. For example, resource allocation may be needed during evacuation to provide support to evacuees in need of assistance (i.e. those who cannot relocate themselves because of being injured or with functional limitations) while minimizing the time to provide such help. The assessment of behavioural travel patterns of population can be used to strategically plan for assisted evacuation. Since in road transport congestion may impact the time to reach a given destination, the knowledge on activities and behavioural patterns of people in need can be helpful to strategically allocate assistance. In addition, in case of use of public transport to aid evacuation, social media data on modal split and behavioural itineraries can be useful to appropriately design the needed public transport resources (e.g. buses). Previous studies focused for instance on the relationship between departure time and evacuation logistics \cite{Chen2022,Deng2021}. 
Many existing studies discuss route decisions \cite{Xie2023,Xu2023,Gao2022,Chen2022,Wachtel2021}. In particular, \cite{Xie2023} looks at disrupted transportation networks and the influence of information availability and stochasticity on evacuation decisions. \cite{Wachtel2021} deals with the planning of evacuation routes for tourists. It highlights how specific populations, such as tourists, may have unique decision-making factors.

\section{Open Problems \label{sec:problems}}
This section discusses the open problems in social media data mining of human behaviour during bushfire evacuations. Research is needed on the open problems to expand the use of social media data in future fire- and hazard-related applications. 

\begin{itemize}
    \item Data Quality and Misinformation: Social media data can be noisy and contain irrelevant or misleading information (including fake news). Distinguishing between reliable information and misinformation is a critical challenge. 
    \item Bias and Representativeness: Social media data may not be representative of the entire population, potentially leading to biased insights. Addressing these biases and ensuring the inclusivity of the analysis is important.
    \item Geolocation Accuracy: Determining the exact location of users during evacuations is crucial. Geolocation data can be inaccurate or missing, making it hard to identify users in affected areas.
    \item Contextual Understanding: Interpreting social media posts requires understanding of the context, as statements may not always be straightforward. Also, determining the urgency and severity of evacuation-related posts can be difficult.
    \item Crisis-Specific Lexicon and Semantics: During bushfire, people may use non-standard language, abbreviations or domain-specific terms. Developing accurate language models and lexicons for crisis-related content can be difficult.
    \item Multimodal Data Analysis: People share information with various media formats, such as text, images and videos. Integrating and analysing these diverse data sources to gain a comprehensive understanding of behaviours is a challenge.
\end{itemize}

\subsection{Data Quality and Misinformation}
Social platforms have become crucial sources of information during emergency events such as bushfire evacuations. However, the abundance of noisy and potentially misleading information on these platforms poses a significant challenge. The prevalence of misinformation can distort behavioral analysis. False reports about fire locations, exaggerated danger assessments, or incorrect evacuation routes may circulate rapidly during crises. {This problem is directly linked to the inherent data limitations of {\bf incompleteness}, {\bf implicitness}, and {\bf low signal-to-noise ratio} discussed in Section \ref{sec:introduction}.} To address this issue, future research on social media data mining of human behaviours during bushfire evacuations can employ various techniques to distinguish between reliable information and misinformation.

\begin{itemize}
    \item Contextual analysis: Valid posts usually include specific details, such as location, time, and personal experiences related to the evacuation process. Data mining algorithms can identify patterns where credible posts consistently share facts with context-rich information. On the contrary, misinformation often lacks these specific details and can contain conflicting information that does not align with the overall context of the event.
    \item Network analysis: When examining connections and interactions between users, data mining algorithms can identify influential and trustworthy sources. Reliable information often comes from users with established networks that have consistently shared accurate updates over time. On the contrary, bots or unreliable sources may exhibit limited connections and sporadic posting patterns. This is particularly important for filtering fake news or other misleading information \cite{Liao2022,Wang20232767}.
    \item Verification mechanisms: During bushfire evacuations, information can change rapidly and rumours can spread quickly. Data mining tools can integrate fact-checking tools to validate the accuracy of information as it emerges. This involves cross-referencing posts with official announcements, news sources, and other reliable channels.
    \item \textcolor{black}{Data Driven} Machine learning: By training models on historical data, these algorithms can learn to recognise characteristics commonly associated with false information. This can include the use of certain keywords, phrasing, or posting patterns. As new posts are analysed, a machine learning model can compare them with learnt patterns to flag potential instances of misinformation.
\end{itemize}

In conclusion, future research on social media data mining of human behaviour during bushfire evacuations needs to effectively address the challenge of distinguishing reliable information from misinformation. Through sentiment analysis, contextual analysis, network analysis, real-time verification, and machine learning models, data mining algorithms can sift through the noise to identify credible sources and accurate information. 

\subsection{Bias and Representativeness}
Addressing biases and ensuring inclusion of all potential evacuees in social media data mining of human behaviours during bushfire evacuations is crucial to obtain accurate and meaningful insights. {This challenge is a direct consequence of the {\bf bias} and {\bf hyperlocal data} limitations inherent to social media, as detailed in the social media discussion in Section \ref{sec:introduction}.} To mitigate bias, future research should study how to effectively employ the following strategies.

\begin{itemize}
    \item Diverse Data Collection: Ensure that the data collected from social platforms cover a wide demographic, geographic and socioeconomic background. This can be done by intentionally seeking out diverse data sources and using techniques like stratified sampling to ensure representation.
    \item Bias Detection and Correction: Implement algorithms and techniques that can identify and quantify biases within the data. By understanding the biases present, researchers can apply correction methods to make the analysis more balanced and representative of the population.
    \item Contextual Understanding: Developing an understanding of the context in which the data were generated is essential. This involves recognising the limitations of social media data, understanding the motivations behind user posts, and accounting for any cultural nuances that might affect the interpretation of the data.
    \item Inclusive Language Models: When analysing text data, researchers use inclusive language models that have been trained to understand and respect diverse perspectives and identities. This can help to accurately capture the nuances of different voices and viewpoints.
    \item Combining Data Sources: To overcome the limitations of a single data source, researchers combine social media with other data sources, such as government reports, surveys, and news articles. This multi-source approach can provide a more holistic understanding of the evacuation behaviours.
\end{itemize}

Future bushfire evacuation-related studies on these strategies can help to address biases and improve the inclusivity of all potential evacuees in the analysis. These studies, in turn, could lead to the retrieval of more accurate information on household evacuation behaviour.

\subsection{Geolocation Accuracy}
During bushfire evacuations, accurate geolocation data is essential to ensure the safety of people. However, relying solely on social media geolocation data can be problematic due to inaccuracies or missing information. {This problem stems from the possibility of {\bf missing or inaccurate location information} as mentioned in the social media discussion in Section \ref{sec:introduction}, and the technical gap of methods that rely on external data sources with {\bf potential runtime overhead}, as discussed in Section \ref{ssec:data_categorization}.} To address this challenge, future studies of social media data mining of evacuation behaviour should find solutions to improve the precision of locating people during evacuations using the following strategies.

\begin{itemize}
    \item Behaviour Pattern Analysis: By analysing user behaviour patterns on social media during bushfire events, valuable information can be gained. These insights can help authorities understand how people respond, where they are likely to gather, and what routes they might take during evacuations. Mining data for keywords, hashtags, and geotags related to the incident can provide real-time information on affected areas and users' movements.
    \item Multi-source Cross-Referencing: By cross-referencing multiple sources of information, including official emergency alerts, news reports, and user-generated content, a more accurate picture of the situation can be developed. This can help verify the accuracy of geolocation data and identify discrepancies. For instance, if a user claims to be in a specific location, but their behaviours suggest otherwise, it raises a red flag for potential inaccuracies.
    \item Geolocation accuracy analysis: Machine learning algorithms can be used to improve the accuracy of geolocation data by predicting users' likely locations based on the behavioural factors found in previous posts, such as travel speed, typical daily routines, and preferred modes of transportation.
\end{itemize}

In summary, future bushfire evacuation-related studies can contribute significantly to addressing the challenge of inaccurate or missing geolocation data. 

\subsection{Contextual Understanding}
Future studies using social media data to understand bushfire evacuations must effectively address the challenge of interpreting the urgency and severity of evacuation-related posts by leveraging contextual insights and user behaviours. {This challenge is exacerbated by the {\bf implicitness and informality} of social media data (Section \ref{sec:introduction}), which makes automated interpretation difficult, as noted in the discussion on data cleaning and categorisation in Sections \ref{ssec:data_cleaning} and \ref{ssec:data_categorization}.} Analysing social media posts with the following techniques can help to distinguish between genuine urgency and misinterpretation.

\begin{itemize}
    \item Hashtag/Keyword Trend Analysis: The frequency and intensity of hashtags/keywords related to evacuation (e.g., "fire," "evacuate," "danger") can indicate the urgency associated with the event. A sudden surge in such terms, coupled with geotags, may suggest an evolving emergency. Conversely, recurring hashtags/phrases in posts could imply routine updates rather than immediate danger.
    \item Sentiment Analysis: Posts infused with fear, desperation, or concern are likely to indicate urgency. However, a balanced sentiment might point to measured updates or general discussions.
    \item User Engagement Analysis: User interactions offer context. Posts with a high volume of reposts, shares, and comments often signify critical information dissemination. In addition, location-based data can help determine whether the poster is actually in the affected area or providing remote information.
    \item Information Flow Analysis: Rapid increase in the frequency of social media in a region can suggest an evolving situation that demands swift action. 
    \item Credibility Assessment: Verified accounts, official emergency response agencies, and reputable news agencies are of greater importance. Cross-referencing with news updates can provide a more accurate understanding. Incorporating machine learning techniques can refine this analysis by learning from historical data and evolving patterns. Training algorithms to recognise these features improves their ability to distinguish between genuine emergencies and non-urgent situations. However, challenges such as sarcasm or misinformation may still persist, making a multidimensional analysis approach crucial.
\end{itemize}

In conclusion, future studies on a holistic approach that combines these techniques can improve contextual understanding with social media data mining of human behaviours during bushfire evacuations. These techniques can help to identify the urgency and severity of evacuation-related posts, thus facilitating a better understanding of household evacuation behaviour.

\subsection{Crisis-Specific Lexicon and Semantics}
Social media can significantly help to understand behaviors during bushfire evacuations, even in the presence of varied language use, such as nonstandard language, abbreviations, and domain-specific terms. {This problem is a direct manifestation of the {\bf informality} of social media (Section \ref{sec:introduction}) and the {\bf rapidly evolving vernacular} of crisis discussions (Sections \ref{ssec:data_collection} and \ref{ssec:data_cleaning}).} Mining these data involves analysing large volumes of user-generated content to extract valuable information. Future studies can integrate the following techniques to address the varied use of language.

\begin{itemize}
    \item Natural Language Processing (NLP): To address the challenge of varied language use, NLP techniques can be employed. This includes training language models, such as GPT, Llama, and Alpaca, on diverse and evolving crisis-related content. These models can learn to decipher context and make sense of nonstandard language, abbreviations, and domain-specific terms by recognising patterns in usage and context.
    \item Sentiment Analysis: Data mining algorithms can identify emergent themes, keywords, and hashtags related to bushfires and evacuations. Sentiment analysis can determine the emotional tone of posts, helping to gauge the urgency and intensity of the situation. Clustering algorithms based on sentiment analysis can group similar posts together for NLP to extract varied language representations about evacuation behaviours in trending communication patterns.
    \item Event Detection: By tracking real-time social activity, authorities can gain insights into evacuation routes, shelter preferences, and emerging obstacles even in the presence of varied language use. This information can be used by NLP algorithms to dynamically extract different representations about emergency response and resources. 
    \item Lexicon Development: A lexicon specific to crisis-related content can be developed using machine learning. This lexicon would include evolving terms, slang, and abbreviations used during emergencies. The training data can be created using NLP, Sentiment Analysis and Event Detection. As the model encounters these terms in context, it can improve its understanding, ensuring an accurate interpretation of posts.
\end{itemize}

In summary, future studies should consider NLP, sentiment analysis, event detection, and lexicon development to decode nonstandard language, abbreviations, and domain-specific terms and in turn, give insights into behaviours during bushfire evacuations.

\subsection{Multimodal Data Interpretation}
The mining of social media data in this context involves extracting valuable information from various media formats to understand critical situations. Future studies must address the challenge of integrating and analysing diverse data sources by employing advanced techniques to merge information and identify behaviours. {This challenge is a significant technical gap, as effectively {\bf integrating multimodal data} (text, images, video) requires sophisticated fusion techniques that go beyond the current state-of-the-art categorization discussed in Section \ref{ssec:data_categorization}.}
\begin{itemize}
    \item Natural Language Processing: Social media platforms host a large number of text-based content, such as posts, reposts, and comments. Natural language processing (NLP) tools can be used to extract sentiment, urgency, and location information from these texts. This helps gauge public sentiment and identify areas that need attention.
    \item Image Analysis: Images shared on social media provide visual evidence of the evolving situation. Image recognition algorithms can identify key elements such as smoke plumes, fire fronts, and damaged infrastructure. These insights enable a better understanding of the behaviours of people and their reasons during bushfires.
    \item Video Processing: Videos offer dynamic information about the crisis. Video analysis tools can detect movement, infer the intensity of the situation based on visual cues like flames and smoke, and even identify landmarks to pinpoint locations. This helps evacuation planners understand emergency situations and environmental factors in bushfires.
    \item Geospatial Data Integration: Integrating geolocation information from social media posts aids in creating maps of evacuation zones and fire extents. Geographic information helps to better connect social media data with location information.

\end{itemize}

In conclusion, related future studies should use techniques such as NLP, image analysis, video processing, and geospatial data integration to enable multimodal data interpretation. These studies can overcome the challenge of handling diverse data sources by extracting valuable information and improving understanding of behaviours, ultimately leading to more efficient emergency planning and public safety.

\section{Conclusions}
Social media analysis for evacuation decision-making has attracted the attention of many researchers, but the use of social media data to investigate human behaviour during bushfire evacuations is still relatively new. This article introduces the concepts and techniques of social media data mining to understand evacuation behaviour during bushfires and discusses its future applications and open problems. It is hoped that, in the near future, these techniques could be helpful for many research projects and applications where a better understanding of evacuation behaviour during bushfires is needed. 

\section{Acknowledgements}

This work was performed under financial assistance award 60NANB22D179 from the U.S. Department of Commerce, National Institute of Standards and Technology. It was also supported by an ARC Future Fellowship by the Australian Government
through the Australian Research Council (Grant No. FT220100618). The authors would like to thank the WUI-NITY team (Jonathan Wahlqvist, Guillermo Rein, Harry Mitchell, Nikolaos Kalogeropoulos, Steve Gwynne, Hui Xie, Peter Thompson, Hamed Mozaffari Maaref, Maxine Berthiaume, Noureddine Bénichou, and Amanda Kimball). We also acknowledge the technical panel of the project for their support and guidance: Carole Adam, Amy Christianson, Tom Cova, Lauren Folk, Abishek Gaur, Paolo Intini, Justice Jones, Bryan Klein, Chris Lautenberger, Ruggiero Lovreglio, Jerry McAdams, Ruddy Mell, Elise Miller-Hooks, Cathy Stephens, Steve Taylor, Sandra Vaiciulyte, Xilei Zhao, Rita Fahy, Lucian Deaton, and Michele Steinberg. Enrico Ronchi also wishes to acknowledge the European Research Council (ERC) for supporting this research through the ERC Consolidator Grant Egressibility (Grant Agreement ID: 101170110).

\bibliography{main}

\end{document}